\documentclass{optica-article}

\journal{opticajournal} 

\articletype{Research Article}

\usepackage{lineno}
\usepackage{amsmath}          
\usepackage{siunitx}          

\begin{document}

\title{Electrical thermography via centimetre-scale fiber-based distributed temperature sensing}

\author{Victor Cochet,\authormark{1} Axel Faccio,\authormark{1} Georgios Stoikos,\authormark{2}, Towsif Taher,\authormark{2} Rob Thew,\authormark{2}, Jérôme Extermann,\authormark{1} and Enrico Pomarico\authormark{2}}

\address{
\authormark{1}HEPIA, HES-SO, University of Applied Sciences and Arts Western Switzerland, Rue de la Prairie 4, 1202 Geneva, Switzerland\\
\authormark{2}Department of Applied Physics, University of Geneva, CH-1211 Geneva, Switzerland
}

\email{\authormark{*}enrico.pomarico@hesge.ch} 


\begin{abstract*} 
We present a Raman-based Distributed Temperature Sensor (RDTS) with centimetre-scale resolution for thermographic analysis of electronic circuits. Temperature is measured along a single-mode fiber routed across a custom printed circuit board (PCB) with \SI{1}{cm^2} heating elements, using optical time-domain reflectometry of Raman signals detected by superconducting nanowire single-photon detectors (SNSPDs).

This approach enables two-dimensional thermal mapping of the PCB under heating configurations with multiple hotspots. A spatial resolution of \SI{3}{cm} and a temperature accuracy of \SI{2}{\celsius} are achieved with an integration time of \SI{5}{minutes}. Thermography can be performed down to \SI{77}{K}, revealing that the PCB’s thermal resistance decreases by nearly an order of magnitude compared to room temperature, due to enhanced convective cooling in liquid nitrogen.

These results establish centimetre-scale RDTS as a robust technique for real-time, spatially resolved thermography of electronic circuits, particularly in regimes where infrared imaging is ineffective, such as at low temperatures or within volumetric electronic architectures.

\end{abstract*}

\section{Introduction}
Electrical thermography refers to the practice of mapping temperature distributions within electrical systems and devices to detect anomalies such as overheating, inefficient components, or incipient faults~\cite{Liu2007}. 
In power distribution systems, this approach helps detect hotspots in transformers, switchgear, and cables under load that may signal failure or fire risk. In electronics manufacturing and testing, thermal imaging can reveal poor solder joints, connector issues, or excessive power dissipation. In large-scale infrastructures, such as solar and wind farms, thermal diagnostics help detect early electrical stress and prevent costly failures ~\cite{Li2024}. 

Infrared (IR) imaging is the dominant method for electrical thermography, offering rapid temperature mapping of electronic circuits and hotspot detection~\cite{Sarawade2018,Balakrishnan2022}. However, its use is restricted to surface measurements, requires careful emissivity calibration, and cannot reveal subsurface or encapsulated heat sources. Moreover, IR thermography becomes ineffective in cryogenic environments, such as quantum computing platforms and superconducting electronic circuits, as cold surfaces emit negligible IR radiation.

To overcome depth penetration and cryogenic limitations, alternative techniques have been developed. Thermoreflectance microscopy offers sub-micron spatial and nanosecond temporal resolution for buried microelectronic layers. Scanning thermal microscopy~\cite{Harnack2025,Zhang2020} and scanning Joule expansion microscopy ~\cite{Qiao2025} achieve nanoscale thermal imaging of individual transistors and interconnects. Fiber Bragg grating (FBG) arrays~\cite{Alhussein2025} and fiber-optic probes enable point-wise temperature sensing, but lack continuous spatial coverage and 3D mapping capability.

Despite their high resolution, these methods are slow, costly, limited in field of view, difficult to scale for volumetric monitoring, and often require contact or invasive installation. These limitations highlight the need for thermal imaging with interior access, centimetre-scale resolution, real-time monitoring, and cryogenic compatibility, a gap that can be addressed by fiber-based distributed temperature sensing (DTS).

DTS utilizes telecom-grade optical fibers and leverages Raman, or Brillouin scattering in combination with time-domain reflectometry to continuously measure temperature along the entire length of a fiber. While traditional DTS provides meter-scale resolution~\cite{Ukil2012,Lu2019}, recent advances using superconducting nanowire single-photon detectors (SNSPDs) have surpassed this limit ~\cite{Dyer2012,Vo2014,Gasser2022}.  In particular, a Raman-based DTS with \SI{3}{cm} resolution over hundreds of meters range has been demonstrated in~\cite{Gasser2022}. In our recent study, we have shown 3D temperature mapping down to approximately \SI{48}{K} using Raman DTS with SNSPDs~\cite{Corradin2024}. These performances have been achieved thanks to SNSPDs high photon detection efficiency, low dark counts, and timing jitter on the order of tens of picoseconds~\cite{Holzman2019}.
By combining centimetre spatial resolution, real-time operation, and cryogenic performance, centimetre-scale DTS offers a unique thermal diagnostic tool. It enables detection of hotspots within circuits, supports dynamic thermal response monitoring, resists electromagnetic interference, and can be deployed across entire boards or 3D modules. 

In this work, we demonstrate the application of a SNSPD-based Raman DTS (RDTS) for thermography of a printed circuit board (PCB) built to simulate electronic circuit thermal anomalies. We first perform heat dissipation measurements to understand the dependence of temperature variations on the electrical current flowing into localized resistors. We then describe the theoretical approach to determine temperature profiles from Anti-Stokes and Stokes Raman signals. After calibration of the DTS using a dedicated heating element, we present the spatial calibration procedure required to generate PCB thermograms. Finally, we report temperature maps of the PCB both at room temperature and at cryogenic temperatures.

\section{Methods}

\subsection{Optical setup}

\begin{figure}[htbp]
\centering\includegraphics[width=9cm]{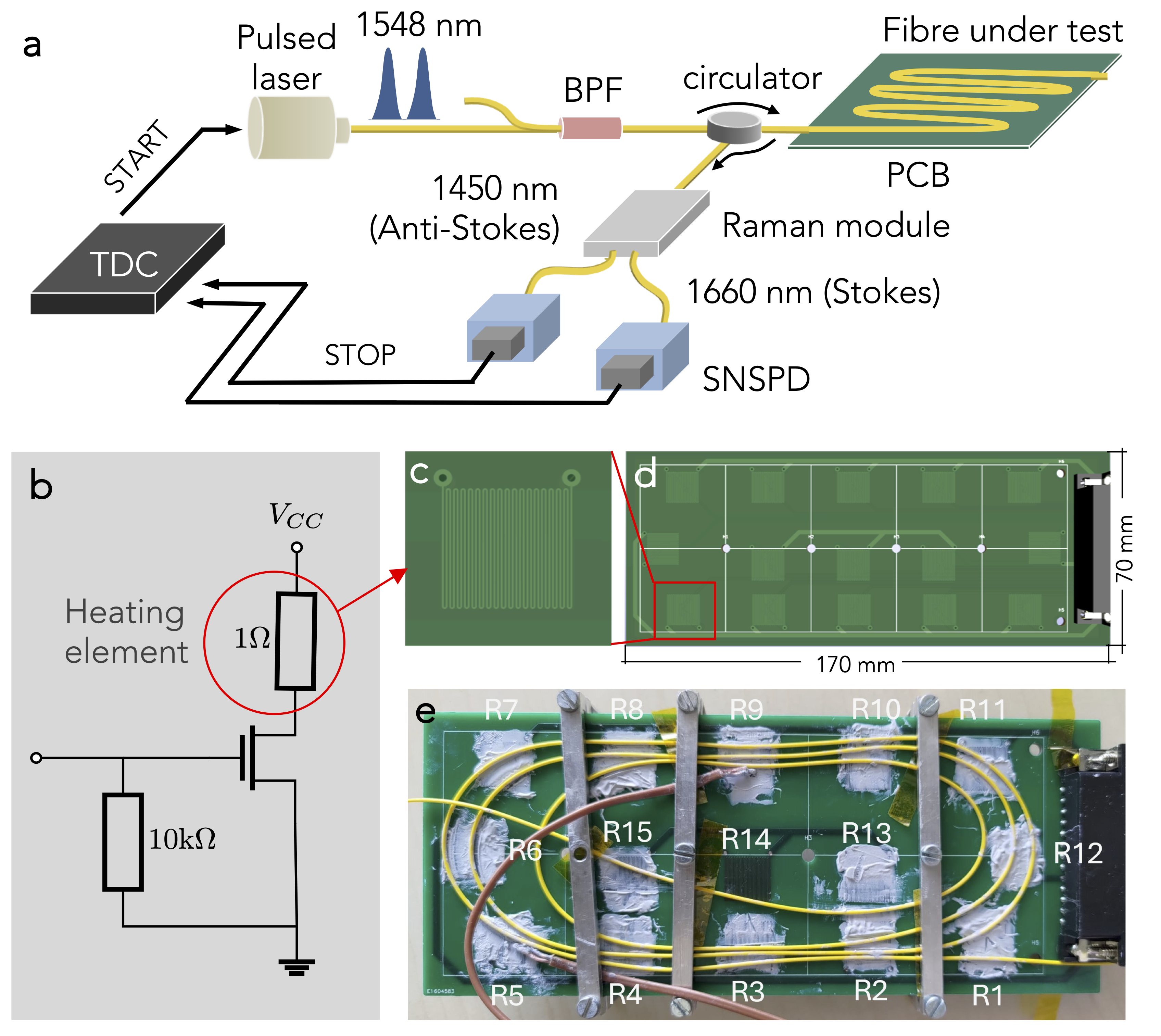}
\caption{Optical setup and printed circuit board (PCB) developed for fiber-based thermography measurements. a) Schematic of the RDTS optical setup. A time-to-digital controller (TDC) triggers a laser emitting pulses at 1548 nm. The pulses pass through a band-pass filter (BPF) and are injected into the fiber under test, which is routed on a custom-designed PCB. A circulator directs the backscattered Raman signal to a wavelength-division multiplexer (WDM) module that separates the AS and S components. These signals are detected by two superconducting nanowire single-photon detectors (SNSPDs), which generate stop signals for the TDC. b) Electrical schematic of the PCB heating elements. c) Footprint of a single heating element on the PCB.
d) Top view of the PCB showing the layout of the 15 heating elements.
e) Photograph of the experimental setup with the fiber positioned on the PCB.}\label{fig:setup}
\end{figure}

The Raman Distributed Temperature Sensor (RDTS) developed in this study utilizes an Optical Time-Domain Reflectometry (OTDR) approach with single-photon counting detectors. A diode laser (PicoQuant, LDH-P-C-N-1550) is triggered by a time-to-digital controller (TDC, ID Quantique, ID900) at a repetition rate of \SI{2.5}{MHz}. It emits pulses at \SI{1548}{nm} with a \SI{1.5}{nm} full width at half maximum (FWHM), \SI{250}{ps} duration, and approximately \SI{30}{mW} peak power (Fig.~\ref{fig:setup}a). These pulses pass through a \SI{13}{nm} band-pass filter to reduce Rayleigh noise before being launched into a single-mode (SM) telecom fiber under test (FUT). A circulator directs the light backscattered from the fiber to a Raman wavelength-division multiplexing (WDM) module (OfLink, RWDMM-456-SM-L-10-FA), which separates the Anti-Stokes (AS) signal at \SI{1450}{nm} (insertion loss: \SI{0.48}{dB}, isolation: \SI{62}{dB}) from the Stokes (S) one at \SI{1660}{nm} (insertion loss: \SI{0.73}{dB}, isolation: \SI{62}{dB}) using a \SI{20}{nm} spectral window. 

Both AS and S signals are detected by superconducting nanowire single-photon detectors (SNSPDs), fabricated from linearly oriented Niobium Titanium Nitride (NbTiN) thin films and housed in a sorption cryostat operating at \SI{0.8}{K}~\cite{Stasi2023}. These SNSPDs exhibit detection efficiencies around 80\%, timing jitter in the range of \SI{30}{ps}--\SI{50}{ps}, and dark count rates below \SI{100}{counts\per\second}. The voltage pulses generated by the SNSPDs upon single-photon detection are amplified and serve as STOP signals for the TDC. 

The FUT is routed along a custom-designed PCB to monitor temperature variations across its surface.

\subsection{Custom-made electronic PCB}
The printed circuit board (PCB) is specifically designed to generate localized heating zones, enabling the simulation of temperature anomalies with controlled heat dissipation. To achieve this, 15 serpentine-shaped conductive tracks, each forming a heating area of approximately \SI{1}{cm^{2}} (Fig.~\ref{fig:setup}c) are embedded at multiple locations across the board (Fig.~\ref{fig:setup}d). The geometry of these tracks is optimized to yield an electrical resistance of about \SI{1}{\ohm} at room temperature. The corresponding electrical schematic of the heating elements is presented in Fig.~\ref{fig:setup}b. 
A MOSFET, selected for reliable operation at cryogenic temperatures, is used to switch the heating elements. A pull-down resistor of \SI{10}{\kilo\ohm} ensures that the heating element remains off when the control signal is inactive. Heating is induced via the Joule effect by driving current through the serpentine resistive tracks. Thermal paste is used on the heating elements to improve contact with the FUT routed on the PCB (Fig.~\ref{fig:setup}e).

\section{Heat dissipation measurements on the PCB}
\begin{figure}[htbp]
\centering\includegraphics[width=12cm]{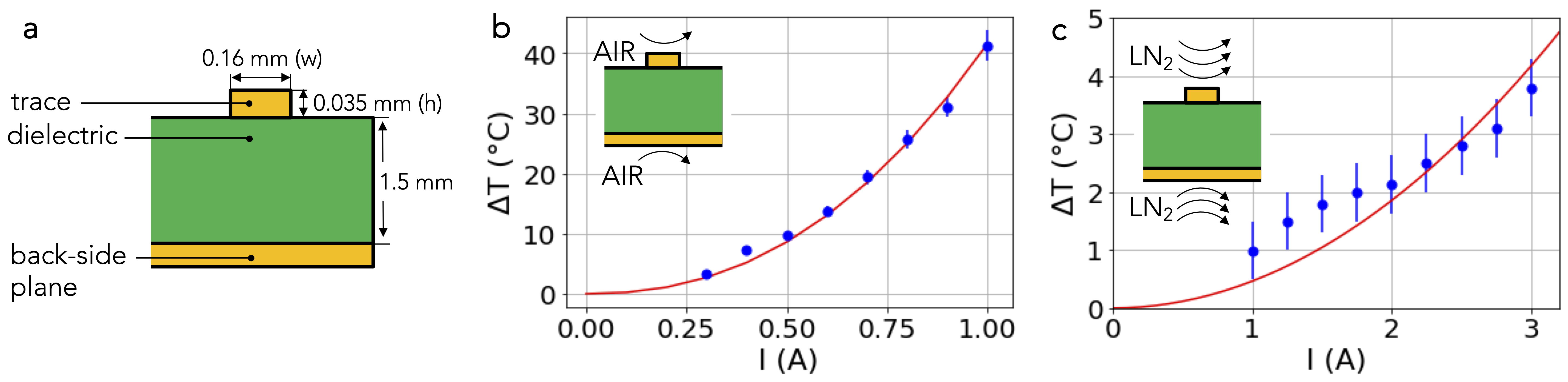}
\caption{Heat dissipation measurements on the PCB. 
    (a) Schematic representation of the PCB layout. 
    (b) Temperature variation on the R9 heating element in ambient air as a function of the electrical current. 
    (c) Temperature variation on the R9 heating element with the PCB immersed in liquid nitrogen (LN$_2$). 
    Quadratic fitting lines are represented in red.
}\label{fig:electrical_measurements}
\end{figure}

Each heating element on the PCB consists of a copper trace with a defined width and thickness, forming a resistive path for the electrical current. The copper trace is deposited on a dielectric substrate, which provides both electrical insulation and mechanical support. On the opposite side, a continuous copper plane serves as the ground or return path (Fig.~\ref{fig:electrical_measurements}a).

A single heating element is characterized at room temperature by an electrical resistance given by
\begin{equation}
R_{\mathrm{el}}(T_0) = \frac{\rho_0 L_0}{hw}, 
\label{eq:rel0}
\end{equation}
where $\rho_0$ is the copper electrical resistivity at room temperature $T_0$, 
$L_0$ is the trace length at room temperature, $h$ its height, and $w$ its width. 
By using $\rho_0 = 1.68 \times 10^{-8}\,\si{\ohm\meter}$ \cite{Lide2004}, and the nominal parameters 
$L_0 = \SI{432}{\milli\meter}$, $w = \SI{0.16}{\milli\meter}$ and 
$h = \SI{35}{\micro\meter}$, we obtain from eq.~\eqref{eq:rel0} 
$R_{\mathrm{el}}(T_0) = \SI{1.3}{\ohm}$.
At temperature $T$, the electrical resistance changes due to variation in copper 
resistivity as well as in trace length because of thermal expansion. 
Assuming linear expansion only along the trace length, the expression of the electrical resistance becomes
\begin{equation}
R_{\mathrm{el}}(T) = \frac{\rho(T) L(T)}{hw} 
= \frac{\rho(T) L_0}{hw}\left[1 + \alpha (T - T_0)\right],
\label{eq:relT}
\end{equation}
where $\rho(T)$ and $L(T)$ are the temperature-dependent copper resistivity and 
trace length, respectively, and $\alpha$ is the copper linear thermal expansion coefficient. 

To simulate temperature anomalies on the PCB, a controlled current is applied to one 
heating element, generating heat through the Joule effect. The electrical power 
dissipated at temperature $T$ can be expressed as
\begin{equation}
P(T) = R_{\mathrm{el}}(T) I^2 = 
\frac{\rho(T) L_0}{hw}\left[1 + \alpha (T - T_0)\right] I^2,
\label{eq:power}
\end{equation}
where $I$ is the current flowing through the heating element. 
The power dissipation causes a temperature rise in the heating element depending on the
thermal resistance $R_{\mathrm{th}}(T)$ of the PCB. Assuming that the temperature variation 
$\Delta T$ is proportional to the dissipated power, eq.~\eqref{eq:power} gives
\begin{equation}
\Delta T = R_{\mathrm{th}}(T)\, P(T) 
= R_{\mathrm{th}}(T)\, R_{\mathrm{el}}(T)\, I^2.
\label{eq:deltaT}
\end{equation}
Eq.~\eqref{eq:deltaT} shows that the temperature rise is expected to scale 
with the square of the current and with the product $R_{\mathrm{th}} (T) \cdot R_{\mathrm{el}}(T)$, 
expressed in $ \si{\celsius\per\ampere\squared}$.

To understand the relation between temperature rise and electrical current, we study heat dissipation corresponding to R9 heating element (cf. Fig.~\ref{fig:setup}e) and measure temperature variations with a thermocouple in contact with it. The PCB is left in ambient air while the current is varied. As shown in Fig.~\ref{fig:electrical_measurements}b, we observe the expected quadratic relation between
$\Delta T$ and $I$ and measure temperature variations of up to \SI{40}{\celsius} for 
currents between 0 and \SI{1}{\ampere}. By fitting the experimental points with a 
quadratic function, we obtain 
$R_{\mathrm{th}}(T_0)\,R_{\mathrm{el}}(T_0) = (40.0 \pm 0.5)\,\si{\celsius\per\ampere\squared}$. 
Using $R_{\mathrm{el}}(T_0) = \SI{1.3}{\ohm}$, this yields 
$R_{\mathrm{th}}(T_0) = \SI{32}{\celsius\per\watt}$.

We then immerse the PCB in liquid nitrogen at \SI{77}{\kelvin} and repeat the 
measurements (Fig.~\ref{fig:electrical_measurements}c). Under these conditions, temperature variations are 
significantly reduced: we observe changes up to only \SI{4}{\celsius}, even with 
currents as high as \SI{3}{\ampere}. The higher required current levels arise from 
the reduction in both electrical and thermal resistances at cryogenic temperatures. 

The reduction of electrical resistance 
at \SI{77}{\kelvin} can be estimated from eq.~\eqref{eq:relT}. 
Taking $\rho(77\,\si{\kelvin}) = 1.5 \times 10^{-9}\,\si{\ohm\meter}$ \cite{Matula1979} and 
$\alpha = 17 \times 10^{-6}\,\si{\per\celsius}$ \cite{Rumble2024}, we obtain 
$R_{\mathrm{el}}(77\,\si{\kelvin}) = \SI{0.12}{\ohm}$. 
This confirms that the drop in electrical resistance is primarily due to reduced resistivity at low temperature. 

The smaller $\Delta T$ values at cryogenic temperatures are not only due to electrical reasons: 
they are also influenced by enhanced convective heat transfer of the PCB in liquid nitrogen. 
Fitting the experimental data with a quadratic function,  
we obtain $R_{\mathrm{th}}(77\,\si{\kelvin}) R_{\mathrm{el}}(77\,\si{\kelvin}) = 
(0.49 \pm 0.04)\,\si{\celsius\per\ampere\squared}$. Using the calculated value 
$R_{\mathrm{el}}(77\,\si{\kelvin}) = \SI{0.12}{\ohm}$, this yields 
$R_{\mathrm{th}}(77\,\si{\kelvin}) \approx \SI{4.1}{\celsius\per\watt}$, 
about 13\% of the value obtained at room temperature.

Assuming convective heat transfer coefficients of 
$h_{\text{air}} \approx \SI{10}{\watt\per\square\meter\per\celsius}$ and 
$h_{N_2} \approx \SI{120}{\watt\per\square\meter\per\celsius}$ for air and liquid nitrogen respectively~\cite{Wang2015}, 
we estimate the thermal resistance ratio as
\[
\frac{R_{\mathrm{th}}(77\,\si{\kelvin}) }{R_{\mathrm{th}}(T_0) } \approx \frac{1/h_{N_2} }{1/h_{\text{air}}} =\frac{10}{120} = 0.083,
\]
indicating that thermal resistance in liquid nitrogen is about 8--9\% of that in air. 
This result is in good agreement with our experimental outcome, especially considering 
that convective coefficients depend strongly on geometry and flow conditions.

\section{Determination of the temperature via the DTS}
In this section, we explain how to determine the temperature at a position $x$ of the FUT routed on the PCB, by measuring the AS and S signals at that position.  
Taking into account the optical phonon population at temperature $T$ described by the Bose--Einstein distribution, the number of photon counts per second at the AS and S wavelengths at position $x$ is given by~\cite{Dyer2012}:
\begin{align}
\mathrm{AS}(T(x)) &= \frac{A}{\exp(C/T(x)) - 1} + N_{\mathrm{AS}}, \label{eq:AS}\\[6pt]
\mathrm{S}(T(x))  &= \frac{B \exp(C/T(x))}{\exp(C/T(x)) - 1} + N_{\mathrm{S}}, \label{eq:S}
\end{align}
where $A$ and $B$ are coefficients including transmission losses through the optical components, detector efficiencies, the peak pump power, as well as the Raman gain coefficient, and $N_{\mathrm{AS}}$ and $N_{\mathrm{S}}$ are the noise rates in the AS and S channels, respectively.  
The constant $C$ is defined as $C = \frac{\hbar \Omega}{k_B}$,
where $\hbar$ is the reduced Planck constant, $\Omega = \SI{13}{\tera\hertz}$ is the spectral shift between the pump and the AS signal (assumed equal for the S signal), and $k_B$ is the Boltzmann constant~\cite{Dyer2012}. In eq.~\eqref{eq:AS}, we assume that the difference in attenuation losses between the AS and S wavelengths is negligible.  

The use of polarization-dependent detectors implies that, even at a fixed temperature, the Raman signals exhibit fluctuations. These variations hinder a direct calibration of the RDTS based solely on the ratio between the AS and S signals at temperature $T$. To overcome this limitation, we consider a reference temperature $T_0$, which we choose to be room temperature, and define the variation in the AS signal with respect to this temperature as
\[
\Delta \mathrm{AS} = \mathrm{AS}(T(x)) - \mathrm{AS}(T_0(x)).
\]
By normalizing this quantity to the S signal (after subtracting its noise $N_S$) and using eq.~\eqref{eq:AS} and ~\eqref{eq:S}, we obtain the following dimensionless expression:
\begin{equation}
\frac{\Delta \mathrm{AS}}{\mathrm{S}(T)-N_S}
= \frac{\mathrm{AS}(T) - \mathrm{AS}(T_0)}{\mathrm{S}(T)-N_S}
= \frac{\mathrm{AS}(T)}{\mathrm{S}(T)-N_S}
 - \frac{\mathrm{AS}(T_0)}{\mathrm{S}(T)-N_S}
\approx C_1 e^{-C/T} + C_2.
\label{eq:ratio}
\end{equation}

In eq.~\eqref{eq:ratio}, the spatial dependence on $x$ has been omitted for clarity and the temperature dependence of the term $\mathrm{AS}(T)/(\mathrm{S}(T)-N_S)$ can be exactly expressed as a function of $T$, while the term $\mathrm{AS}(T_0)/(\mathrm{S}(T)-N_S)$ is approximated as a constant $C_2$.  
This simplification is justified since $\mathrm{AS}(T_0)$ is fixed, while $\mathrm{S}(T)$ varies only weakly with $T$. Indeed, we estimate that a temperature increase of \SI{50}{\kelvin} above $T_0=\SI{296}{\kelvin}$ leads to a maximum deviation of about 5\% in $C_2$, whereas for $T_0=\SI{77}{\kelvin}$ the variation remains below 1\%.

With these approximations, the RDTS calibration requires only two constants. By isolating the temperature $T$ from eq.~\eqref{eq:ratio}, we obtain the following expression for the DTS-derived temperature at position $x$:
\begin{equation}
T_{\mathrm{DTS}}(x) 
= -\frac{\hbar \Omega}{k_B \ln \left[ 
\frac{1}{C_1}\left(
\frac{\mathrm{AS}(x) - \mathrm{AS}(T_0(x))}{\mathrm{S}(x)-N_S} - C_2
\right) \right]}.
\label{eq:T_dts}
\end{equation}

Here, the subscript "DTS"' indicates that the temperature value is obtained from the RDTS measurement.  

\section{Results}
\subsection{Temperature calibration at room temperature}
\begin{figure}[htbp]
\centering\includegraphics[width=12cm]{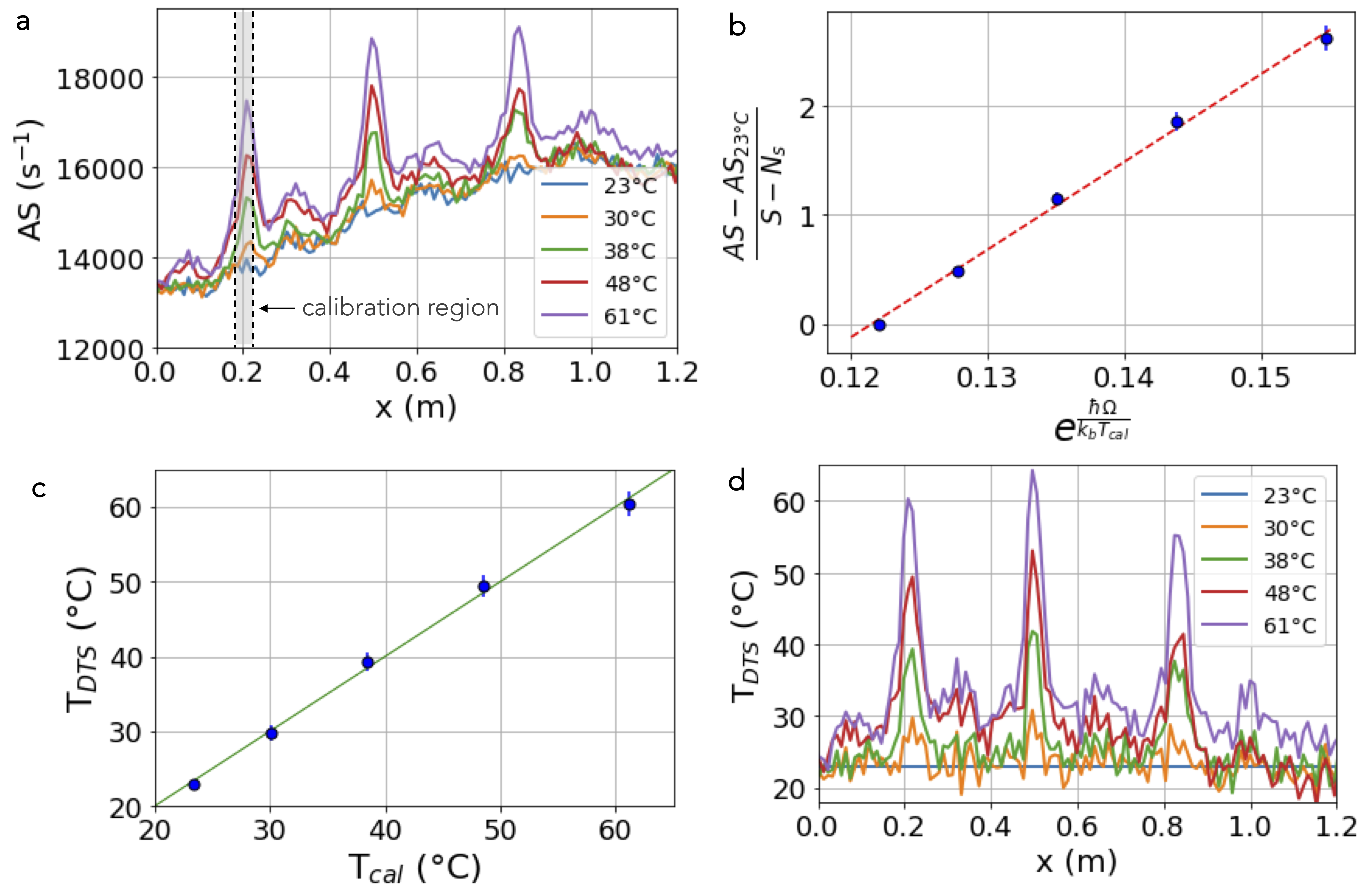}
\caption{Temperature calibration of the RDTS in ambient air. 
    (a) AS signal as a function of position $x$ along the FUT, measured for different heating temperatures of the R9 element on the PCB. The grey-shaded area indicates the region used for calibration.
    (b) Ratio of the AS signal (after subtraction of its value at \SI{23}{\celsius}) to the Stokes signal (after subtraction of the dark count noise) within the calibration region, plotted as a function of $\exp(\hbar \Omega / (k_B T_{\mathrm{cal}}))$. The fitting red line is used to extract the calibration constants. 
    (c) DTS-estimated temperature $T_{\mathrm{DTS}}$ as a function of the calibrated thermocouple temperature $T_{\mathrm{cal}}$. The green line represents the ideal one-to-one correspondence. 
    (d) Temperature profiles estimated by the DTS along the FUT for different temperatures in the calibration region.
}\label{fig:calibration_room_temperature}
\end{figure}

To calibrate the RDTS, we use a calibrated thermocouple placed in contact with the R9 heating element (cf. Fig.~\ref{fig:setup}e).  
Figure~\ref{fig:calibration_room_temperature}a shows the AS signal measured along the FUT for various temperatures ranging from \SI{23}{\celsius} to \SI{61}{\celsius}.  

At \SI{23}{\celsius}, the signal varies with the position $x$ along the fiber. This behavior arises from the polarization dependence of the SNSPD detector. Indeed, the polarization of the optical pulses changes during propagation through the fiber. Since the Raman gain depends on polarization, the intensity of the scattered light also varies with position along the FUT, thereby generating a modulated signal on the detector~\cite{Dyer2012}.  

The three peaks observed in Fig.~\ref{fig:calibration_room_temperature}a correspond to the three passages of the FUT across the R9 heating element.  
To calibrate the thermometer, we average the AS signal in the grey-shaded section of the OTDR trace in Fig.~\ref{fig:calibration_room_temperature}a, which corresponds to the thermocouple position shown in Fig.~\ref{fig:setup}a.  
As explained in the previous section, we calculate the variation of the AS signal with respect to its value at \SI{23}{\celsius} in the calibration region (not shown in Fig.~\ref{fig:calibration_room_temperature}):
\[
\Delta \mathrm{AS} = \mathrm{AS}(T(x)) - \mathrm{AS}(\SI{23}{\celsius}).
\]
We then calculate the ratio of $\Delta \mathrm{AS}$ over the Stokes signal corresponding to the grey-shaded region. In Fig.~\ref{fig:calibration_room_temperature}b, this normalized quantity is plotted as a function of $\exp(\hbar \Omega / k_B T)$.  
According to eq.~(7), we perform a linear fit to obtain the calibration constants:
\[
C_1 = 81 \pm 3, \qquad C_2 = -9.8 \pm 0.4.
\]

Fig.~\ref{fig:calibration_room_temperature}c shows the good agreement between the estimated DTS temperatures $T_{\mathrm{DTS}}$ and the calibrated thermocouple temperatures $T_{\mathrm{cal}}$. By error propagation, the uncertainties on $T_{\mathrm{DTS}}$ are found to lie between \SI{1}{\celsius} and \SI{2}{\celsius}.  

Finally, temperature profiles for all calibration points are computed across all fiber positions $x$, as shown in Fig.~\ref{fig:calibration_room_temperature}d. The profile at \SI{23}{\celsius} is flat, as this temperature is used as reference for calculating temperature variations.  

\subsection{Spatial calibration}
\begin{figure}[htbp]
\centering\includegraphics[width=10cm]{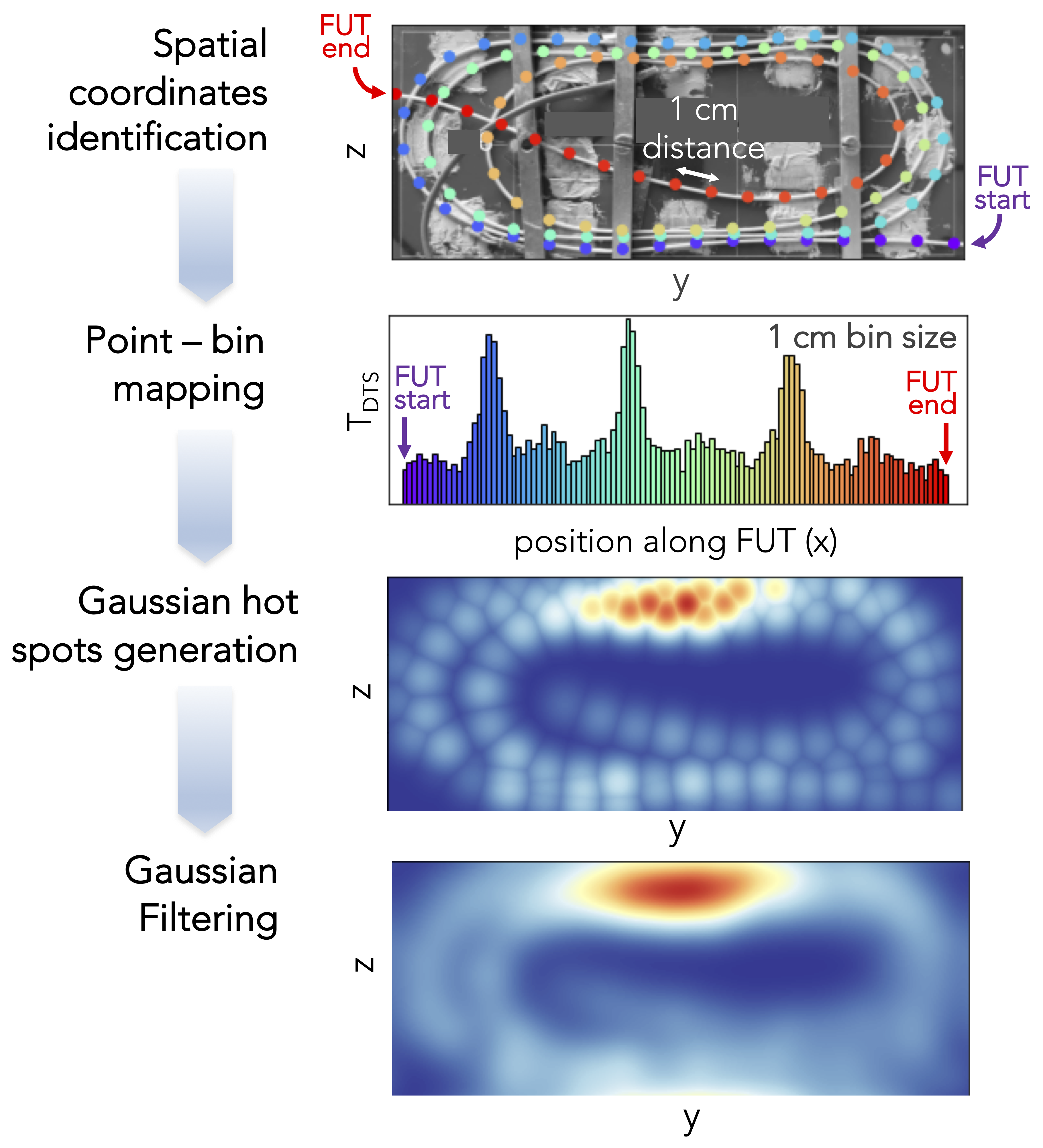}
\caption{Spatial calibration for PCB fiber-based thermography. Spatial coordinates of 1~cm distant points selected along the FUT are first identified (top). Each point is then mapped to a specific 1~cm bin in the temperature profile. A color code is used to distinguish the points (middle-top). A Gaussian hot spot with 1~cm FWHM is generated at each selected position (middle-bottom). Gaussian filtering is then applied to smooth the resulting temperature map (bottom).
}\label{fig:spatial_calibration}
\end{figure}

To produce thermograms of the PCB, one should map 1~cm bins of the temperature profiles to specific spatial regions on the PCB surface. The reconstructed temperature values are then used to produce a temperature color map. The pipeline used to perform this spatial calibration, represented schematically in Fig.~\ref{fig:spatial_calibration}, is based on the following steps: 

\begin{enumerate}
    \item \textbf{Spatial Coordinates Identification:} Distinct points along the FUT’s path are selected at regular 1~cm intervals, as shown in the top-right image of Fig.~\ref{fig:spatial_calibration}. By fixing a Cartesian reference system, spatial coordinates of the selected points along the FUT are identified. Coordinates can be found either by digitizing the image of the PCB or by fixing the FUT starting point and computing mathematically the next coordinates. In the top-right image of Fig.~\ref{fig:spatial_calibration}, each point is assigned a unique color to aid in visual tracking across the next steps.
    
    \item \textbf{Point--Bin Mapping:} Spatial points are then mapped to corresponding 1~cm bins in the temperature profile (middle-top plot in Fig.~\ref{fig:spatial_calibration}). The temperature profile is represented as a histogram of $T_{\mathrm{DTS}}$ as a function of the position along the FUT. The color coding allows to make a direct visual link between physical positions on the PCB and temperature values or bins on the histogram.
    
    \item \textbf{Gaussian Hot Spot Generation:} At the position of each mapped point, a localized Gaussian hot spot is generated on a 2D temperature map (middle-bottom image of Fig.~\ref{fig:spatial_calibration}). Each Gaussian function has a full width at half maximum (FWHM) of 1~cm, reflecting the spatial extension of the chosen portions of fiber.
    
    \item \textbf{Gaussian Filtering:} A 2D Gaussian smoothing filter with a FWHM of 1~cm is applied to the obtained temperature map (bottom image of Fig.~\ref{fig:spatial_calibration}), resulting in a continuous and visually interpretable thermal image of the PCB.
\end{enumerate}

\subsection{PCB fiber-based thermography at room temperature}
\begin{figure}[htbp]
\centering\includegraphics[width=12cm]{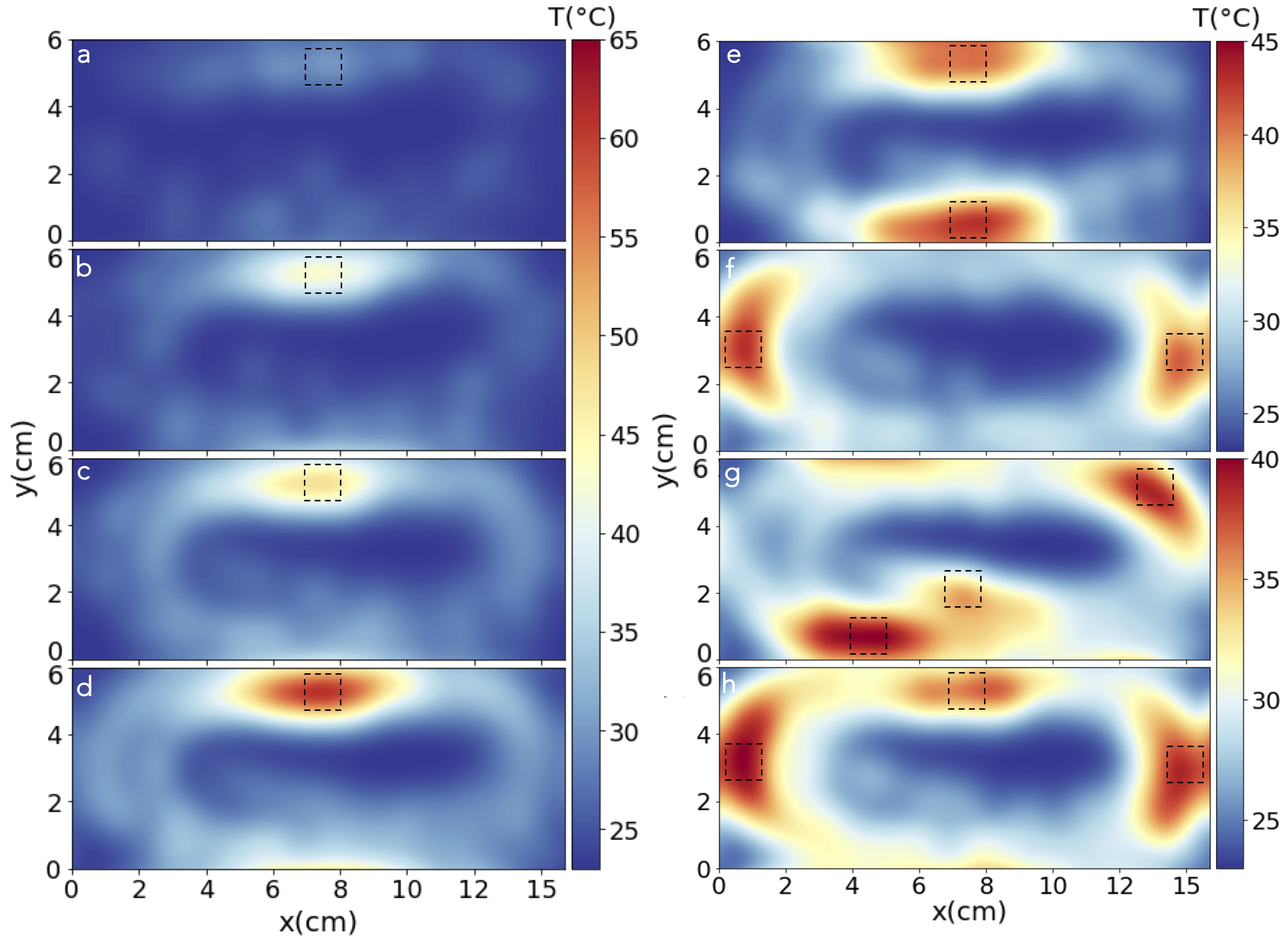}
\caption{PCB thermography under different spatial and temperature conditions. Temperature maps for different heating temperatures on the R9 element (a-d), on R9 - R3 (e), on R6 - R12 (f), on R4-R11-R14 (g), and on R6-R9-R12 elements (h). The dashed squares indicate the positions of the respective heating elements.
}\label{fig:thermography_room_temperature}
\end{figure}

We first implement our technique at room temperature, by activating one to three distinct heating elements on the PCB in various spatial configurations. Fig.~\ref{fig:thermography_room_temperature} presents distinct thermal maps of the PCB, with each subfigure (a--h) corresponding to a specific heating configuration. The color scale indicates the measured temperature, while the dashed boxes highlight the positions of the heating elements, serving as benchmarks for evaluating localization accuracy.  

Fig.~\ref{fig:thermography_room_temperature}a--d show thermograms recorded by progressively heating the R9 resistor. As the applied current increases, the temperature at R9 rises, reaching up to $65~^\circ\mathrm{C}$. Initially, the map appears nearly uniform with only a slight elevation at the R9 location (Fig.~\ref{fig:thermography_room_temperature}a). A localized hotspot begins to form (Fig.~\ref{fig:thermography_room_temperature}b). The heating becomes then more prominent, reaching approximately $50~^\circ\mathrm{C}$ (Fig.~\ref{fig:thermography_room_temperature}c). Finally, a well-defined hotspot exceeding $60~^\circ\mathrm{C}$ appears (Fig.~\ref{fig:thermography_room_temperature}d). Heat spreads laterally from the source, with a spatial extent of around 3~cm, which is mainly determined by the spatial resolution of our DTS system~\cite{Gasser2022}. The thermogram in Fig.~\ref{fig:thermography_room_temperature}d also shows partial heating across the PCB, suggesting thermal diffusion via conduction through the device. The obtained thermograms are consistent with the temperature variations measured by electronic measurements, as discussed in Section~3.  

Fig.~\ref{fig:thermography_room_temperature}e--h display configurations where multiple heating elements are simultaneously activated. In these cases, the temperature peaks at around $45~^\circ\mathrm{C}$ due to the lower current delivered to each individual heater, as the total current is distributed across the activated elements.  
In Fig.~\ref{fig:thermography_room_temperature}e, two hotspots appear at the top and bottom of the image, corresponding to activated elements R9 and R3. In Fig.~\ref{fig:thermography_room_temperature}f, heat is localized in the left and right corners, corresponding to R6 and R12. Fig.~\ref{fig:thermography_room_temperature}g shows three distinct sources: R11, R14, and R4. In Fig.~\ref{fig:thermography_room_temperature}h, a different triplet (R6, R9, and R12) is activated, producing three thermal hotspots.  

Together, these results confirm that centimetre-scale DTS can successfully resolve multiple heat sources over a planar electronic system, capture complex spatial thermal distributions, and retain sensitivity across a broad temperature range. 

\subsection{PCB fiber-based thermography under liquid nitrogen cooling}
In this section, we describe the implementation of our technique for PCB thermography at cryogenic temperatures. The PCB is immersed in liquid nitrogen, and the same calibration procedure described in Section~5.1 is performed, using the calibrated thermocouple placed in contact with the R9 heating element (Fig.~\ref{fig:setup}e).  

\begin{figure}[htbp]
\centering\includegraphics[width=12cm]{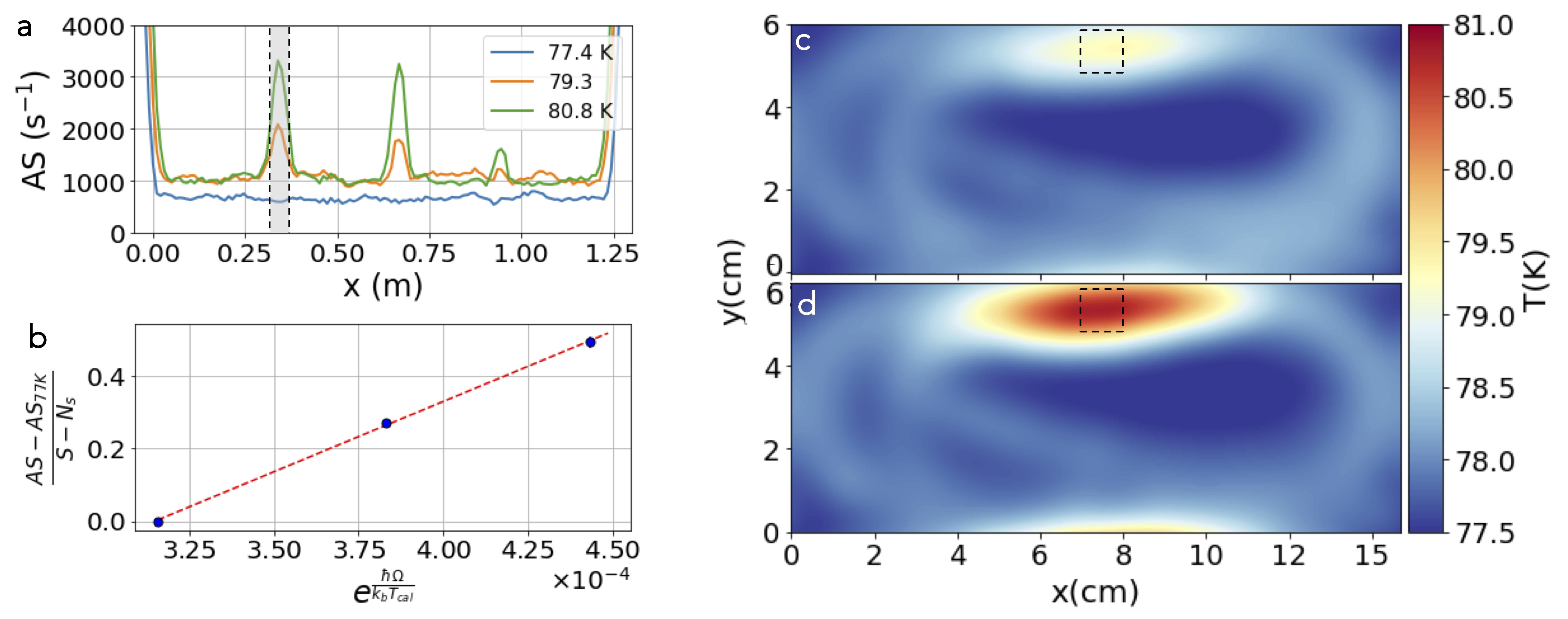}
\caption{Temperature calibration of the RDTS and PCB fiber-based thermography in liquid nitrogen. 
    (a) AS signal as a function of position along the FUT, measured for different heating temperatures of the R9 element on the PCB. The grey-shaded region corresponds to that used for calibration. 
    (b) Ratio of the AS signal (with its value at 77~K subtracted) and the $S$ signal (with its dark count noise subtracted) within the calibration region, plotted as a function of $e^{\hbar \Omega / (k_B T_{\mathrm{cal}})}$. The fitting line is used to extract the calibration constants. 
    (c--d) Temperature maps for different heating temperatures on the R9 element.
}\label{fig:calibration_cryo}
\end{figure}

Fig.~\ref{fig:calibration_cryo}a shows the AS signal measured along the FUT for temperatures ranging from 77 to $\SI{81}{K}$. As explained in Section 3, the temperature variations of the PCB resistors are limited by the decrease in electrical resistance of the heating elements and by the enhanced convective heat dissipation in liquid nitrogen, which is stronger than in air.
As for room temperature measurements, we observe three peaks, resulting from the FUT passing three times by the heated element. To calibrate the thermometer under these conditions, we average the signals in the grey region in Fig.~\ref{fig:calibration_cryo}a and calculate the quantity  
\[
\frac{AS(T) - AS(77~\mathrm{K})}{S - N_S}.
\]  
Fig.~\ref{fig:calibration_cryo}b shows this quantity as a function of $e^{\hbar \Omega / (k_B T)}$. Linear fitting provides the calibration constants  
\[
C_1 = (384 \pm 9) \cdot 10^2, \quad C_2 = -1.21 \pm 0.03.
\]  

Fig.~\ref{fig:calibration_cryo}c and~\ref{fig:calibration_cryo}d show the temperature maps for different heating conditions of the R9 element. These results demonstrate the ability of our technique to localize temperature variations on the order of 1~K under cryogenic conditions, which is not possible with conventional infrared thermography methods.

\section{Conclusion}
In this work, we demonstrated the thermography of electronic circuits using Raman-based distributed temperature sensing. Our system achieves a spatial resolution of \SI{3}{cm} and a temperature accuracy of $\SI{2}{\celsius}$ with an integration time of 5 minutes.

We first analyzed the temperature variations on the PCB arising from electrical power dissipation. These variations are determined by both the electrical resistance of the copper traces forming the heating elements and the PCB’s thermal resistance. At cryogenic temperatures, the reduced temperature rise is attributed to the lower electrical resistivity of copper and the decreased thermal resistance due to enhanced convective heat transfer in liquid nitrogen, which drops from \SI{32}{\kelvin\per\watt} at room temperature to \SI{4.1}{\kelvin\per\watt} at 77 K.

We then reconstructed two-dimensional thermograms of the PCB under different heating configurations at room temperature, demonstrating the ability of our system to resolve simultaneous hotspots within a 6 cm $\times$ 15 cm area. At cryogenic conditions down to 77 K, the system successfully detected centimetre-scale temperature variations as small as 1 K, thereby operating in environments inaccessible to infrared thermography.

The spatial resolution of this technique is limited by the duration of the optical pulses, approximately \SI{250}{ps}. Using shorter pulses, one can exploit the very low timing jitter of SNSPDs to achieve spatial resolution below \SI{1}{cm}.

Fiber-based distributed temperature sensing employing SNSPDs thus provides an effective, non-invasive, and real-time diagnostic tool for identifying and localizing thermal anomalies in electronic modules with centimetre-scale resolution. Beyond PCBs, this thermography technique is well suited for densely integrated or volumetric electronic systems, particle detectors, cryogenic infrastructures and superconducting devices, where conventional thermography is limited.

\section{Back matter}

\begin{backmatter}
\bmsection{Funding}
This work was supported by the University of Applied Sciences and Arts Western Switzerland (HES-SO) and the Swiss State Secretariat for Research and Innovation (SERI) (Contract No. UeM019-3).

\bmsection{Acknowledgment}
 We would like to acknowledge Félix Bussières and Lorenzo Stasi for insightful discussions.

\bmsection{Disclosures}
The authors declare no conflicts of interest.

\bmsection{Data Availability Statement}
The data that support the findings of this study are available from the corresponding author upon reasonable request.

\end{backmatter}


\end{document}